\documentclass[12pt]{article}
\pdfoutput=1
\usepackage{graphicx}
\usepackage{amssymb}
\usepackage{amsmath}
\usepackage{srcltx}
\usepackage{color}
\usepackage{cite}
\usepackage{url}
\usepackage{ntheorem}
\usepackage{psfrag}
\usepackage[bookmarks,colorlinks=false]{hyperref}

\newcommand{\Sing}{\mathrm{Sing}}

\newtheorem{Theorem} {\sc  Theorem\rm} [section]

\newtheorem{Lemma} [Theorem] {\sc  Lemma\rm}
\newtheorem{Proposition} [Theorem] {\sc  Proposition\rm}

\theorembodyfont{\upshape}

\theoremsymbol{\ensuremath{\diamondsuit}}
\newcommand{\fcoco}{\small}
\theorembodyfont{\fcoco}\theoremseparator{}\theoremindent0.5cm

\theoremstyle{nonumberplain}\theorembodyfont{\fcoco}
\theoremseparator{}\theoremindent0.5cm

\DeclareFontFamily{OT1}{rsfs}{}
\DeclareFontShape{OT1}{rsfs}{m}{n}{ <-7> rsfs5 <7-10> rsfs7 <10-> rsfs10}{}
\DeclareMathAlphabet{\mycal}{OT1}{rsfs}{m}{n}

{\catcode `\@=11 \global\let\AddToReset=\@addtoreset}
\AddToReset{equation}{section}

\newcounter{mnotecount}[section]

\AddToReset{figure}{section}
\renewcommand{\themnotecount}{\thesection.\arabic{mnotecount}}

\newcommand{\mnotex}[1]
{\protect{\stepcounter{mnotecount}}$^{\mbox{\footnotesize
$
\bullet$\themnotecount}}$ \marginpar{
\raggedright\tiny\em
$\!\!\!\!\!\!\,\bullet$\themnotecount: #1} }

\newcommand{\ptccheck}[1]{\mnote{ptcchecked:#1}}
\renewcommand{\ptccheck}[1]{\mnote{\checkmark (ptc #1)}}

\newcommand{\jlcax}[1]{}
\newcommand{\eean}{\nonumber\end{eqnarray}}

\newcommand{\kk}[1]{}

\newcommand{\beq}{\begin{equation}}

\newcommand{\FS}       
                  {F}

\newcommand{\HS} 
       {H_{\mbox{\scriptsize volume}}}

{\ptc{this should be removed in the oberwolfach version}}%

\newcommand{\eeal}[1]{\label{#1}\end{eqnarray}}
\newcommand{\bed}{\begin{deqarr}}
\newcommand{\eed}{\end{deqarr}}
\newcommand{\bedl}[1]{\begin{deqarr}\label{#1}}
\newcommand{\eedl}[2]{\arrlabel{#1}\label{#2}\end{deqarr}}

\newcommand{\mcU}{{\mycal U}}

\newcommand{\bel}[1]{\begin{equation}\label{#1}}
\newcommand{\bea}{\begin{eqnarray}}
\newcommand{\bean}{\begin{eqnarray}\nonumber}
\newcommand{\beal}[1]{\begin{eqnarray}\label{#1}}
\newcommand{\eea}{\end{eqnarray}}

\newcommand{\qed}{\hfill $\Box$ \medskip}
\newcommand{\proof}{\noindent {\sc Proof:\ }}
\newcommand{\be}{\begin{equation}}
\newcommand{\eeq}{\end{equation}}
\newcommand{\ee}{\end{equation}}
\newcommand{\beqa}{\begin{eqnarray}}
\newcommand{\eeqa}{\end{eqnarray}}
\newcommand{\beqan}{\begin{eqnarray*}}
\newcommand{\eeqan}{\end{eqnarray*}}
\newcommand{\ba}{\begin{array}}
\newcommand{\ea}{\end{array}}

\newcommand{\mnote}[1]
{\protect{\stepcounter{mnotecount}}$^{\mbox{\footnotesize
$
\bullet$\themnotecount}}$ \marginpar{
\raggedright\tiny\em
$\!\!\!\!\!\!\,\bullet$\themnotecount: #1} }

\newcommand{\warn}[1]
{\protect{\stepcounter{mnotecount}}$^{\mbox{\footnotesize
$
\bullet$\themnotecount}}$ \marginpar{
\raggedright\tiny\em
$\!\!\!\!\!\!\,\bullet$\themnotecount: {\bf Warning:} #1} }

\newcommand{\R}{\mathbb R}

\newcommand{\ptc}[1]{\mnote{{\bf ptc:}#1}}

\newcommand{\beqar}{\begin{deqarr}}
\newcommand{\eeqar}{\end{deqarr}}

\newcommand{\beaa}{\begin{eqnarray*}}
\newcommand{\eeaa}{\end{eqnarray*}}

\renewcommand{\ptccheck}[1]{}
\begin{document}
\title{On the structure of the ergosurface of Pomeransky-Senkov black rings}
\author{Julien Cortier \\ Institut de Math\'ematiques et de
Mod\'elisation de Montpellier\\ Universit\'e Montpellier 2
}
\maketitle{}
\date{}
\begin{abstract}
We study the properties of the ergosurface of the Pomeransky-Senkov black rings,
and show that it splits into an ``inner'' and an ``outer'' region.
As for the singular set, the topology of the ``outer ergosurface'' depends upon the value of parameters.
\end{abstract}

\tableofcontents

\section{Introduction}
\label{intro}
There has been recently a lot of interest in studying higher-dimensional solutions of the vacuum Einstein equations,
especially since the discovery, in 2001 by Emparan and Reall in~\cite{EmparanReall}, of five-dimensional black hole space-times in vacuum,
whose sections of the event horizon are not homeomorphic to a sphere.
Those space-times are usually referred to as ``black rings'', since the topology of
these sections is $S ^1 \times S ^2$ instead.

In this work, we focus on a particular feature, the
\emph{ergosurface}, of a family of solutions discovered in 2006
by Pomeransky and Senkov in~\cite{PS}, also known as ``doubly
spinning black rings''. The Emparan-Reall space-times, also
called ``singly-spinning black rings'', can be seen as a
limiting case of this family (see, e.g., the appendix
of~\cite{CCG}). It has been shown in~\cite{EmparanReall} that
they possess an ergosurface homeomorphic to $\R \times S^1
\times S^2$. In the remainder of this paper, we first recall
some definitions and facts about the Pomeransky-Senkov family
of space-times (Section~\ref{PS}) and establish the notations.
These follow the material introduced in~\cite{CCG}. Then we
study in details the ergosurface (Section~\ref{ergo}), which
turns out to split into two subsets, the first one being the
``usual'' ergosurface lying outside the black hole, the second
lying ``under'' all the Killing horizons. We finish by proving
the results pointed out in~\cite{Durkee} about the existence of
two distinct regimes for the topological nature of the
``upper'' ergosurface.

\section{Pomeransky-Senkov black rings}
\label{PS}
Here we follow the notations and conventions introduced in~\cite{CCG}.
The Pomeransky-Senkov family of metrics is:

\begin{eqnarray}
ds^2 &= & \frac{2 H(x,y)k^2}
{(1-\nu)^2(x-y)^2}\left(\frac{dx^2}{G(x)}-
\frac{dy^2}{G(y)}\right)  -2\frac{ J(x,y)}{H(y,x)}d\varphi d\psi\nonumber\\
&&-
\frac{H(y,x)}{H(x,y)}(dt+\Omega)^2-\frac{F(x,y)}{H(y,x)}d\psi^2+
\frac{F(y,x)}{H(y,x)}d\varphi^2\; ,
\label{eq:line-element1}
\end{eqnarray}
where
\begin{eqnarray*}
H(x,y)=\lambda ^2+2 \nu  \left(1-x^2\right) y \lambda +2
   x \left(1-\nu ^2 y^2\right) \lambda -\nu ^2+\nu
   \left(-\lambda ^2-\nu ^2+1\right) x^2 y^2+1,\\
F(x,y)=\frac{2 k^2}{(x-y)^2 (1-\nu)^2}
\left(\left(1-y^2\right) \left(\left((1-\nu
   )^2-\lambda ^2\right) (\nu +1)+\right.\right.\\
\left.\left.+y \lambda  \left(-\lambda ^2-3
   \nu ^2+2 \nu +1\right)\right) G(x)+\right.\\
\left.+\left(-(1-\nu ) \nu
   \left(\lambda ^2+\nu ^2-1\right) x^4+\lambda  \left(2 \nu
   ^3-3 \nu ^2-\lambda ^2+1\right) x^3+\left((1-\nu )^2-\lambda
   ^2\right) (\nu +1) x^2+\right.\right.\\
\left.\left.+\lambda  \left(\lambda ^2+(1-\nu
   )^2\right) x+2 \lambda ^2\right) G(y)\right),\\
J(x,y)=\frac{2 k^2 \left(1-x^2\right) \left(1-y^2\right)
   \lambda  \sqrt{\nu } \left(\lambda ^2+2 (x+y) \nu  \lambda
   -\nu ^2-x y \nu  \left(-\lambda ^2-\nu
   ^2+1\right)+1\right)}{(x-y) (1-\nu )^2},\\
G(x)=\left(1-x^2\right) \left(\nu  x^2+\lambda  x+1\right)
 \;,
\end{eqnarray*}
and where $\Omega$ is a 1-form given by
$$
\Omega=M(x,y)d\psi+P(x,y)d\varphi
 \;,
$$
with
\begin{eqnarray*}
M(x,y)
&=& \frac{2 k\lambda\sqrt{(\nu +1)^2-\lambda
   ^2} (y+1) (-\lambda +\nu -2 \nu  x+\nu  x ((\lambda
   +\nu -1) x+2) y-1)}{(1-\lambda +\nu ) H(y,x)}
   \\
&=:& \frac{\sqrt{(\nu +1)^2-\lambda
   ^2}\hat M(x,y)}{ (1-\lambda +\nu )  H(y,x)}
\\
P(x,y)
 &= &\frac{2k\lambda  \sqrt{\nu } \sqrt{(\nu
   +1)^2-\lambda ^2} \left(x^2-1\right) y}{H(y,x)}
\\
 &=: &\frac{2\sqrt \nu \hat P(x,y)}{H(y,x)}
   \;,
\end{eqnarray*}
where $\hat P$ and $\hat M$ are polynomials in all variables.

The parameter $k$ is assumed to be in $\R^*$, while the
parameters $\lambda$ and $\nu$ have been restricted
in~\cite{PS} to belong to the set%
\footnote{Strictly speaking, $\nu=0$ is allowed in~\cite{PS}.
It is shown there that this corresponds to Emparan-Reall
metrics (compare Appendix in~\cite{CCG} ), which have already been
analysed elsewhere~\cite{ChCo}, and so we only consider
$\nu>0$.}
\bel{3IX.1}  \mcU:=\{(\nu,\lambda):\ \nu\in (0,1)\;,\
2\sqrt{\nu}\le \lambda< 1 +\nu\}
  \;.
\ee
The coordinates $x$, $y$, $\phi$,
$\psi$, $t$ vary within the ranges $-1\leq x\leq 1$,
$-\infty<y<-1$, $0\leq\phi\leq 2\pi$, $0\leq\psi\leq 2\pi$ and
$-\infty<t<\infty$.

We also introduce the quantities
$$
y_h:= -\frac{\lambda-\sqrt{\lambda^2-4\nu}}{2\nu} \;, \quad y_c:=
-\frac{\lambda+\sqrt{\lambda^2-4\nu}}{2\nu} \;,
$$
In~\cite{CCG}, for every $(\nu,\lambda) \in \mcU$, the submanifold delimited by the coordinates above, with the additional condition $y_h < y < -1$, and equipped with the metric~(\ref{eq:line-element1}), has been shown to be a regular asymptotically flat space-time, with the asymptotically flat region located at $(x,y)=(-1,-1)$. In fact, the set $\{y=y_h\}$ corresponds to an event horizon, $\{y=y_c\}$ to another Killing horizon, and the space-time has been shown to extend smoothly through these horizons up to the singular set $\Sing:=\{H(x,y)=0\}$, where the curvature tensor blows up at least for some values of the parameters.
At this stage, it is convenient to work with a new coordinate $Y$, defined as
$$
Y=-\frac{1}{y}\,,
$$
because the singular set has components which blow up to
infinity in $(x,y)$-coordinates.
%
Since $H(x,y)$ is a second-order polynomial in $y$, we can now
write $\Sing$ as the union of the graphs of the functions
$y_+(x)$ and $y_-(x)$, or, better, we write $\Sing$ in
$(x,Y)$-coordinates as the union of the graphs of $Y_+(x)$ and
$Y_-(x)$, both defined on a subset of $[-1,1]$ such that the
discriminant of the polynomial $H(x,y)$ is non-negative.
Equivalently, if we define $\hat H(x,Y):= Y^2 H(x,-1/Y)$, then
$Y_{\pm}(x)$ are the roots of the second order polynomial $\hat
H(x,Y)$ in $Y$.

In fact, we can observe in these coordinates that the analytic extension done so far up to $\Sing$ can be performed further across the set $\{Y=0\}$,
since the branches $x \mapsto Y_\pm(x)$ are not defined on the whole set $[-1,1]$, but only on a subinterval of $(-1,1]$.
Following the notations of~\cite{CCG}, we have
the inequalities (wherever $Y_+$ and $Y_-$ are defined):
$$
Y_+(x) \leq Y_-(x) < Y_c < Y_h\,.
$$
These confirm that the singular set $\Sing$ lies ``under'' both Killing horizons.
On the other hand, the singular set has the remarkable property to have a varying topology, depending on the values of $(\nu,\lambda) \in \mcU$.
Indeed, the topology is $\mathbb R \times \mathbb T^3$ when $\nu + \lambda < 1$, $\mathbb R \times S^1 \times S^2$ when $\nu + \lambda > 1$, and
a ``pinched'' $\mathbb R \times S^1 \times S^2$ when $\nu + \lambda = 1$. These results are illustrated on Figure 1.3 of~\cite{CCG}.
Some other properties of the singular set have been derived (see Theorem 5.6 in~\cite{CCG}).
Most of them come from the study of the polynomial $\tilde W$, which is, up to a factor $\frac14$,
the discriminant of the second-order polynomial in $x$, $H(x,y)$:
\bel{tildew}
\tilde W:= \nu  \left(\left(x^2-1\right)^2 \nu
   \lambda ^2+x \left(\lambda ^2+2 x \lambda -\nu
   ^2+1\right) \left(2 \lambda  \nu +x \left(\lambda
   ^2+\nu ^2-1\right)\right)\right)
 \;.
\ee
In particular we recall the following result, which turns out to be useful for the next section:
\begin{Lemma}
 \label{L8.23}
In the region $\{-1\leq x \leq 0\;,\;y<y_c\}$, the two branches
$y_\pm$ which exist for small negative values of $x$ meet smoothly
at some $\bar{x}\in (-1,0)$, where $\bar{x}$ is a simple root of
$\tilde{W}$.
\end{Lemma}
We observed in~\cite{CCG}, (see Remark 5.3) that the root $\bar x$ is simple, and is the largest negative one $\tilde W$, and therefore
 the function $(\nu,\lambda) \mapsto \bar x (\nu,\lambda)$ defined in $\mcU$ is continuous.
\section{The ergosurface}
 \label{ergo}

The ergosurface of the Pomeransky-Senkov space-time is the set of zeros of
$$
g_{tt}=-\frac{H(y,x)}{H(x,y)}\;.
$$
Since this function cannot vanish on the singular set $\Sing$
for allowed values of the parameters and variables (see
Section 3 of~\cite{CCG}), the ergosurface coincides with the set of
zeros of $H(y,x)$. In this section we derive a more precise
version of the results of~\cite{Durkee} about the topology of
the ergosurface,
introduced below. Similarly to the singular set $\{H(x,y)=0\}$,
this topology depends upon the values of the parameters $\nu$
and $\lambda$. In what follows, we prove that the ergosurface
exists and splits into two parts which extend over all the
region $x \in [-1,1]$ (Proposition~\ref{ergoexist}). Next,
Proposition~\ref{P21I.1} shows that one of these parts,
referred to as the {\em outer} ergosurface, lies above the
event horizon $\{y=y_h\}$, while Proposition~\ref{P21I.1-}
shows that the other part, referred to as the {\em inner}
ergosurface, lies under both Killing horizons. Finally Theorem
\ref{theo:ersurface-topology} discusses the topology of the
ergosurface for the allowed values of $\lambda$ and $\nu$.

We start by showing that the ergosurface is globally defined:

\begin{Proposition}\label{ergoexist}
The equation $H(y,x)=0$ has solutions $y_{e\pm}(x)$ in
$\mathbb{R}\cup \{\infty\}$ for all $x$ in $[-1,1]$.
\end{Proposition}

\proof
In order to see this, we begin by noting the identity
\begin{equation}
\label{HyxtoHxy}
H(y,x)=H\left(\nu x,\frac y\nu\right)\;,
\end{equation}
holding for any  $\nu\ne 0$, and which enables us to use some
of the properties of the singular set $\Sing$ recalled in Section~\ref{PS}.
Hence, it is sufficient to show that the polynomial
$\tilde W(x)$, defined by the formula~(\ref{tildew}) above, is nonnegative for $x \in
[-\nu,\nu]$. To do so, we write
\ptccheck{12XI}
\bel{Wpnu}\tilde W(x=\nu)=\nu^2(\lambda^2 + 2\nu \lambda + \nu^3 - \nu)(\lambda^2\nu + 1 + 2\nu \lambda - \nu^2)\;,
\ee
\bel{Wmnu}
\tilde W(x=-\nu)=\nu^2(\lambda^2 - 2\nu \lambda + \nu^3 - \nu)(\lambda^2\nu + 1 - 2\nu \lambda - \nu^2)\;.\ee
We need two lemmata:

\begin{Lemma}\label{ergoexist1}
$\tilde W(x=\nu)$ and $\tilde W(x=-\nu)$ are positive.
\end{Lemma}

\noindent{\sc Proof of the lemma~\ref{ergoexist1}.}   For~(\ref{Wpnu}),
the second factor $(\lambda^2 + 2\nu \lambda + \nu^3 - \nu)$ is
positive since we have $\lambda^2\geq 4\nu > \nu$ for
admissible $\nu$ and $\lambda$, and the third factor
$(\lambda^2\nu + 1 + 2\nu \lambda - \nu^2)$ is obviously
positive as well there. The proof of the positivity
in~(\ref{Wmnu}) requires some more effort. Let us begin with
its second factor,
$$h_1(\nu,\lambda)=\lambda^2 - 2\nu \lambda + \nu^3 - \nu.$$
We have $\partial_\lambda h_1=2(\lambda - \nu ) >0$, hence
$\lambda \mapsto h_1(\nu,\lambda)$ is increasing on
$[2\sqrt{\nu},1+\nu)$, and since
\ptccheck{12XI}
$$h_1(\nu,2\sqrt{\nu})=\nu(3-4\sqrt{\nu}+\nu^2)=\nu(1-\sqrt{\nu})^2(3+2\sqrt{\nu}+\nu)\;,$$
positive for all $\nu$ in $(0,1)$, therefore $h_1$ is positive on $\mcU$.
Then, for the third factor of~(\ref{Wmnu}) :
$$h_2(\nu,\lambda)=\lambda^2\nu - 2\nu \lambda + 1 - \nu^2\;,$$
%

 The first derivative reads
$\partial_\lambda h_2=2\nu(\lambda-1)$, hence the minimum on
the interval $[2\sqrt{\nu},1+\nu)$ of $h_2(\nu,\cdot)$ is
reached at $\lambda = \lambda_m := \max(1,2\sqrt{\nu})$. From
the definition above, $\lambda_m=1$ for $\nu \in (0,1/4]$
(critical point), and $\lambda_m =2\sqrt{\nu}$ for $\nu \in
[1/4,1)$ ($h_2(\nu,\cdot)$ is increasing in the whole interval
$[2\sqrt{\nu},1+\nu)$ in this case).
But on the one hand, we have $h_2(\nu,1)=1-\nu-\nu^2$ positive
for all $\nu$ in $(0,1/4]$, and on the other hand,
\ptccheck{12XI}
$$h_2(\nu,2\sqrt{\nu})=3\nu^2-4\nu\sqrt{\nu}+1=(1-\sqrt{\nu})^2(3\nu+2\sqrt{\nu}+1)\;,$$
positive in particular on $[1/4,1)$. This finishes to show that
all the factors of~(\ref{Wmnu}) are positive, and hence to
prove the lemma.
\qed

\bigskip

Next, we prove a second lemma:
\begin{Lemma}\label{ergoexist2}
For the particular values $(\nu,\lambda=1-\nu)$ of
$(\nu,\lambda)$ in $\mcU$, the whole interval $[-\nu,\nu]$ is
included in the set $\tilde \Omega_{\nu,\lambda}=\{x:\tilde
W(x) \ge 0\}$, and $\tilde W(x)$ has four real valued roots.
\end{Lemma}

\noindent{\sc Proof of the lemma~\ref{ergoexist2}.}   In this
case (which corresponds to $\nu \in (0\ ,\ 3-2\sqrt{2})$),
$\tilde W(x)$ takes the form
$$\tilde W(x)=\nu^2 (1-\nu)^2(x-1)(x+1)(x-(\sqrt{5}+2))(x+\sqrt{5}-2)\;.$$
Here, all the factors but the last one of $\tilde W$ are positive for
$x$ in $[-\nu_0,\nu_0]$, where $\nu_0:=3-2\sqrt{2}<1$.  In
order to determine the sign of the last factor $x+\sqrt{5}-2$, we see that for
all $x$ in $[-\nu_0,\nu_0]$, we have
$$x+\sqrt{5}-2 \geq -\nu_0 + \sqrt{5}-2 = \sqrt{5}+2\sqrt{2}-5\;,$$
the last term being positive, as  can be seen from the
inequality
$$(\sqrt{5}+2\sqrt{2})^2 = 13+4\sqrt{10}>13+4\sqrt{9}=5^2
 \;,
$$
and Lemma~\ref{ergoexist2} is proved.
\qed

These two lemmata enable us to conclude the proof of
Proposition \ref{ergoexist}: Indeed, recall, from the
Lemma~\ref{L8.23} and the remark that follows at the end of Section~\ref{PS}, that the map $\bar x$ is continuous
on $\mcU$, and so is the map $(\nu,\lambda) \mapsto \bar x
(\nu,\lambda) + \nu$. 
The connectedness of
$\mcU$ implies that its image by the map $\bar x + \nu$ is a
connected subset of $\mathbb R$ and hence an interval. Clearly
$0$ cannot belong to this interval because
$\tilde{W}(x=-\nu) > 0$ as shown in Lemma~\ref{ergoexist1} and
thus, either
$$\bar x (\nu,\lambda) < -\nu\ ,\ \forall (\nu,\lambda) \in \mcU\ ,$$
or
$$\bar x(\nu,\lambda) > -\nu \ ,\ \forall (\nu,\lambda) \in \mcU\ .$$
But for $\lambda=1-\nu$, since $\bar x$ is negative by
definition, Lemma~\ref{ergoexist2} entails $\bar x(\nu,1-\nu) <
-\nu$ and hence only the possibility $\bar x (\nu,\lambda) <
-\nu$ for all $(\nu,\lambda)\in \mcU$ remains.

Then,
assume by contradiction that there exists $(\nu_q,\lambda_q)$ in $\mcU$
and $x_q$ in $[-\nu_q,\nu_q]$ such that $\tilde W (x_q)$ is
negative. By Lemma~\ref{ergoexist1}, $\tilde W (x = \pm \nu_q)$ are
both positive and therefore there is an even number of roots
(counting multiplicity) of $\tilde W(x)$ inside the interval
$[-\nu_q,\nu_q]$, and at least two of them or distinct. Since
there cannot be four roots in $[-\nu_q,\nu_q]$ (as $\bar x$ is
outside this interval), there are exactly two distinct ones in
$[-\nu_q,\nu_q]$.
This implies that, for
$(\nu,\lambda)=(\nu_q,\lambda_q)$, the polynomial $\tilde W
(x)$ has four distinct roots; two are in the interval
$[-\nu_q,\nu_q]$, one in the interval $(-\infty,\nu_q)$ and,
given that $\tilde{W}(x)$ is positive for $|x|$ large enough,
the remaining root must lie also in the interval
$(-\infty,-\nu_q)$. Therefore $(\nu_q,\lambda_q)$ belongs to the
connected subset $\mcU_1 =\mcU\cap \{\lambda < \chi
(\nu)\}$, which was defined in Lemma 5.1 of~\cite{CCG} as the domain where $\tilde W$ has four distinct roots,
and so does $(\nu,1-\nu)$ according to Lemma~\ref{ergoexist2}. The roots of
$\tilde W (x)$, for values of $(\nu,\lambda)$ in $\mcU_1$, are
simple and thus they are smooth functions of $(\nu,\lambda)$ in
$\mcU_1$.

Let us denote by $x_4(\nu,\lambda)$ the largest of
these roots. Again, $x_4(\mcU_1)\subset\mathbb R$ is a real
interval as $\mcU_1$ is connected and $x_4$ is continuous.

The previous considerations imply that the greatest root of
$\tilde{W}(x)$ for $(\nu,\lambda)=(\nu_q,\lambda_q)$ fulfills
$x_4(\nu_q,\lambda_q)< \nu_q$ and from Lemma~\ref{ergoexist2}
we deduce $x_4(\nu,1-\nu)=2+\sqrt{5} > \nu\ ,\ \forall \nu \in
(0,1)$. Hence, the continuous map $(\nu,\lambda) \mapsto x_4(\nu,\lambda) - \nu$ changes sign on $\mcU_1$.
Therefore, by the intermediate value theorem, there exits some $(\nu,\lambda) \in \mcU_1$ such that
$x_4(\nu,\lambda) = \nu$. But this is impossible since $\tilde W(x_4(\nu,\lambda))=0$ by definition of $x_4$, whereas $\tilde W(\nu) > 0$ as shown in
Lemma~\ref{ergoexist1}.

This shows that $\tilde W(x) \geq 0$, for $x$ in
$[-\nu,\nu]$, and the corresponding solutions $y_\pm(x)$ are
therefore defined on $[-\nu,\nu]$, except $y_+(x)$ which diverges
at $x=0$. However, $Y_\pm$ are both well defined on $[-\nu,\nu]$. The relation~(\ref{HyxtoHxy}) finishes the proof.
\qed

We have just shown that $H(y,x)=0$ has two solutions
$y_{e-}(x)$, $y_{e+}(x)$ which are defined $\forall x\in
[0,1]$ (allowing an infinite value for $y_+$ at $x=0$). Note that, from~(\ref{HyxtoHxy}), they are related to the solutions $y_{\pm}(x)$ of $H(x,y)$ by
\begin{equation}
y_{e\pm}(x)=\nu y_{\pm}(\nu x)\;,
\label{eq:yetoypm}
\end{equation}
for all $x$ in $[-1,1]$. Equivalently, we have in $(x,Y)$-coordinates:
\begin{equation}
Y_{e\pm}(x)=\frac{1}{\nu} Y_{\pm}(\nu x)\;,
\label{eq:YetoYpm}
\end{equation}
which means that, in the $(x,Y)$-coordinates, the graphs of $Y_{e\pm}$ are obtained from the graphs of $Y_\pm$ by
an homothety of center $(0,0)$, and of ratio $1/\nu$.
In particular, we have
$Y_{e-}(x) > Y_{e+}(x)$, and we define the \emph{outer
ergosurface} as the set $\{y=y_{e-}(x)\;:\; x \in [-1,1]\}$, or, in $(x,Y)$-coordinates, $\{Y=Y_{e-}(x)\;:\; x \in [-1,1]\}$.
We now prove that this last set lies above the event horizon. To
that end we need a lemma:

\begin{Lemma}
 \label{L21I.1}
  $x\mapsto y_{e-}(x)$ reaches its minimum at one of the ends
$\pm 1$.
\end{Lemma}

\begin{figure}[h]
\vspace{-1cm}
\begin{center}
 \includegraphics[width=0.7\textwidth,bb=0 0 290 287]{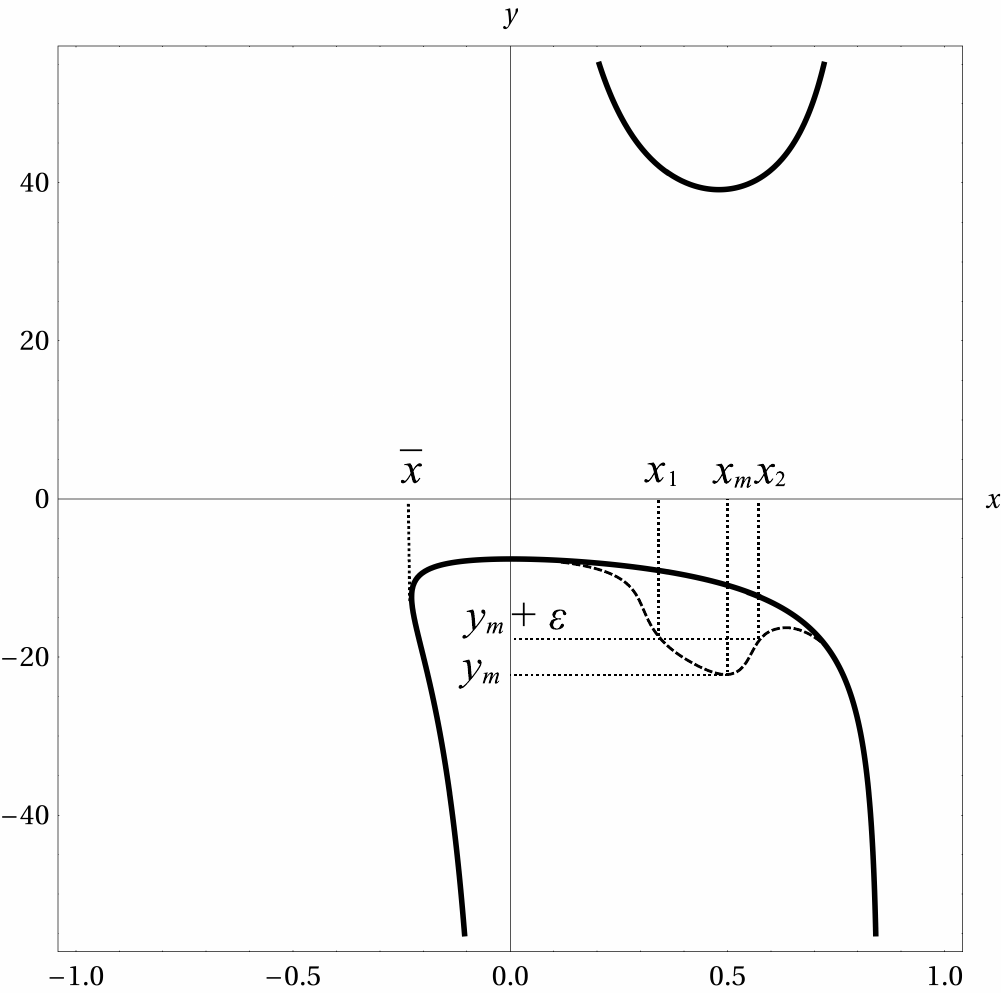}
\caption{Depiction of the geometric set-up used in the proof of lemma \ref{L21I.1}.}
\label{minye-}
\end{center}
\end{figure}

\proof   From the equality $y_{e-}(x)=\nu y_{-}(\nu x)\;,$ it
suffices to prove that the function $x\mapsto y_-(x)$ reaches
its minimum at one of the ends $-\nu$ or $\nu$ in the interval
$[-\nu,\nu]$. If not, there would exist a strict local minimum
of $y_-$ located at $x=x_m \in (-\nu,\nu)$. Let
$y_m:=y_-(x_m)$. Then, for $\varepsilon >0$ small enough, there
exist two values $x_1$ and $x_2$ such that
$y_-(x_i)=y_m+\varepsilon$, with $-\nu < x_1 < x_m < x_2 <
\nu$, and such that $y_-$ is strictly decreasing in a
neighbourhood of $x_1$ (see figure~\ref{minye-}). Furthermore, recall that the graphs of
$y_+$ and $y_-$ join smoothly at $x=\bar x$, where $\bar x$
(introduced in Lemma~\ref{L8.23}) is the greatest negative root
of $\tilde W$ (in particular we have $-1 < \bar x <
-\nu$).
Hence, in the interval $[\bar x,x_1)$, the graphs of $y_+$ and $y_-$ form a smooth connected
curve $\mathcal{C}$ (if $x_1 >0$, there is another connected
component of the graph of $y_+$ which is not of interest here).
Therefore the points of $\mathcal{C}$ can be represented by
pairs $(x(s),y(s))$ where $x(s)$, $y(s)$ are smooth functions
in the interval $(0,1)$ fulfilling the relations $x(0)=x_1$,
$x(s)\in [\bar x,x_1)$, $\forall s\in (0,1)$ and
\begin{equation}
y_-(x(s))=y(s),\; s\in (0,\bar s),\quad
y_+(x(s))=y(s),\; s\in (\bar s,1),
\label{eq:c-curve}
\end{equation}
where $x(\bar s)=\bar x$. Also if $x_1<0$, one knows from
subsection 5.4 of~\cite{CCG} that $y_+(x)<y_-(x)$ for any
$x\in(\bar x, x_1]$, and if $x_1\geq 0$, we have $y_+(x)
\rightarrow -\infty$ as $x \rightarrow 0^-$. Thus, from
(\ref{eq:c-curve}) it follows that there exists a real number
$p\in (0,1)$ such that $y(p)< y_-(x_1)=y(0)$ and $p>\bar s$. On
the other hand the property $\partial_x y_-(x_1) <0$ implies
that there exists a real number $q\in (0,1)$, $q<\bar s$ such
that $\bar x \leq x(q) < x_1$ and $y_-(x(q))>y_-(x_1)$ which
according to (\ref{eq:c-curve}) is equivalent to $y(q)>y(0)$.
Hence we have the inequalities $y(p)<y(0)<y(q)$ with $q<p$ and
given that $y(s)$ is continuous in the interval $[q,p]$ the
intermediate value theorem tells us that there exists a real
number $s^*\in [p,q]$ such that $y(s^*)=y(0)$. If we define
$x^*\equiv x(s^*)$, then eq. (\ref{eq:c-curve}) implies that,
either $y_{-}(x^*)=y(0)=y_m+\varepsilon$ if $s^*<\bar s$ or
$y_+(x^*)=y(0)=y_m+\varepsilon$ if $s^*>\bar s$. In any case
the conclusion is that the numbers $x^* < x_1 <x_2$ are three
distinct roots of the polynomial $H(x,y_m+\varepsilon)$ of
order two in $x$, which is impossible.
\qed

\begin{Proposition}
 \label{P21I.1}
We have the inequality $y_{e-}(x)>y_h$, for all $x$ in $[-1,1]$.
\end{Proposition}

\proof We first write the expressions:
$$H(y_h,-1)=\frac{\lambda
(1+\nu-\lambda)}{2\nu}\left(\left(\lambda-\sqrt{\lambda^2-4\nu}\right)(\lambda+\nu-1)-4\nu\right)\:,$$
$$H(y_h,1)=-\frac{\lambda(1+\nu+\lambda)}{2\nu}
\left(\left(\lambda-\sqrt{\lambda^2-4\nu}\right)(\lambda+1-\nu)-4\nu\right)\:.$$
We will check in a moment that they are both negative. This
will imply the proposition~\ref{P21I.1}. In order to see this,
note first that the negativity of these expressions, together
with the positivity of both $H(-1,\pm 1)$ (see Section~3
of~\cite{CCG}) show that both (continuous) functions $y \mapsto
H(y,\pm 1)$ vanish on the interval $(y_h,-1)$. And since
$$\lim_{y \rightarrow -\infty} H(y,\pm 1) = + \infty\ ,$$
these two functions vanish once again in the interval
$(-\infty,y_h)$. Hence, we obtain that $y_{e-}(x=\pm 1)$
belongs to the interval $(y_h,-1)$, and $y_{e+}(x=\pm 1)$
belongs to the interval $(-\infty,y_h)$. 
This entails $y_h<y_{e-}(x=\pm 1)$ and from Lemma~\ref{L21I.1}
we conclude that $y_h<y_{e-}(x)$, $\forall x\in [-1,1]$ as
desired. We now turn to the proof of the negativity of the
quantities $H(y_h,\pm 1)$ written above: $H(y_h,-1)$ has the
sign of the second factor,
$\left(\lambda-\sqrt{\lambda^2-4\nu}\right)(\lambda+\nu-1)-4\nu\;.$
But we have the inequalities
$0<\lambda-\sqrt{\lambda^2-4\nu}<\lambda<2$ and
$\lambda+\nu-1<2\nu$ for admissible $\nu$ and $\lambda$, so the
negativity of $H(y_h,-1)$ follows. Then, $H(y_h,1)$ has the
sign of $-u(\nu,\lambda)$, where
$$u(\nu,\lambda)=\left(\lambda-\sqrt{\lambda^2-4\nu}\right)(\lambda+1-\nu)-4\nu\;.$$
With the change of variables $\lambda=2\sqrt{\nu}\cosh(\eta)$, we get
$$u(\nu,\lambda)=2\sqrt{\nu}\left(\sqrt{\nu}e^{-2\eta}+(1-\nu)e^{-\eta}-\sqrt{\nu}\right)\;;$$
the roots of the right-hand term, which is an order-two polynomial
in $e^{-\eta}$, are $-1/\sqrt{\nu}$ and $\sqrt{\nu}$. But the
following equivalence holds:
$$e^{-\eta}>\sqrt{\nu}\  \Leftrightarrow\  \lambda < 1+\nu\;,$$
thus $u(\nu,\lambda)$ is positive for any admissible
$(\nu,\lambda)$.

\qed

We now turn to the inner part of the ergosurface
$$
 \{y=y_{e+}(x)\;:\; x \in [-1,1]\}
 \;,
$$
or, in $(x,Y)$-coordinates,
$$
 \{Y=Y_{e+}(x)\;:\; x \in [-1,1]\}
 \;.
$$
We wish to prove that the inner ergosurface lies beyond the Killing
horizons, and even beyond the singular set $\{H(x,y)=0\}$, in the
sense that
$$
Y_{e+} \leq Y_+ \leq Y_- < Y_c < Y_h
$$
wherever these functions are defined. This is illustrated in Figures~1.3
and~1.4 of~\cite{CCG}. 
This inner part reaches and crosses the set $\{y=\pm\infty\}=\{Y=0\}$ at $x=0$ only, 
we therefore make use of the coordinates $(x,Y)$ in the following:
\begin{Proposition}\label{P21I.1-}
We have the inequality $Y_{e+}(x)<Y_c$, for all $x$ in $[-1,1]$.
Moreover, we have $Y_{e+}(x)\leq Y_+(x)$ for all $x \in \tilde
\Omega_{\nu,\lambda}$, with equality only at $x=0$.
\end{Proposition}
\proof Let us first write:
$$H(y_c,-1)=\frac{\lambda
(1+\nu-\lambda)}{2\nu}\left(\left(\lambda+\sqrt{\lambda^2-4\nu}\right)(\lambda+\nu-1)-4\nu\right)\:,$$
$$H(y_c,1)=-\frac{\lambda(1+\nu+\lambda)}{2\nu}
\left(\left(\lambda+\sqrt{\lambda^2-4\nu}\right)(\lambda+1-\nu)-4\nu\right)\:.$$
Again, these expressions are both negative. Indeed, we have the
inequalities
$\lambda+\sqrt{\lambda^2-4\nu} < 1+\nu + (1-\nu) =2$ for
allowed values of $\lambda$ and $\nu$. Since $\lambda + \nu - 1
< 2\nu $, the whole factor
$\left(\lambda+\sqrt{\lambda^2-4\nu}\right)
(\lambda+1-\nu)-4\nu$ is negative, thus $H(y_c,-1)$ is
negative. Then, it is straightforward to see that
$H(y_c,1)<H(y_h,1)$, itself negative, when comparing both
expressions. Since we have the limits $H(y,\pm 1) \rightarrow +
\infty$ as $y \rightarrow -\infty$, the function 
$H(y,\pm 1)$ must have a zero in the interval $(-\infty, y_c)$ and therefore we are
led to the inequality $y_{e+}(x=\pm 1)<y_c$ \footnote{Recall
from the proof of proposition \ref{P21I.1} the inequality
$y_h<y_{e-}(x=\pm 1)$ which rules out the possibility
$y_{e-}(x=\pm 1)<y_c$.} which in turn entails $Y_{e+}(\pm 1) <
Y_c$. Now, we have that the function $Y_{e+}(x)$ does not admit
any strict maximum on the interval $(-1,1)$ (the proof of this
fact is similar to the proof of Lemma~\ref{L21I.1}) from which
we conclude that $Y_{e+}(x) < Y_c\ ,\ \forall x \in [-1,1]$.

In order to show the second part, we first compare $Y_{e+}$ and
$Y_+$ in a neighbourhood of $x=0$. To do so, we write the
asymptotic expansion:
$$Y_+(x)=-\nu x
\left(1-\frac{1-\nu}{2\nu\lambda}((1+\nu)^2-\lambda^2)x +
o(x)\right)\:;$$
so that, using the fact that $Y_{e+}(x)=Y_+(\nu x)/\nu$, we obtain
$$
 Y_{e+}(x)-Y_+(x)=-\frac{(1-\nu)^2((1+\nu)^2-\lambda^2)}{2\nu\lambda}x^2
 + o(x^2)
 \:.
$$
The
order-two term in the above expansion is negative for
allowed values of the parameters $\nu$ and $\lambda$.
Therefore, we get $Y_{e+}\leq Y_+$ in a neighbourhood of $x=0$,
with equality at $x=0$ only: $Y_{e+}(0)=Y_+(0)=0$. Furthermore,
this inequality extends globally to the set $\tilde
\Omega_{\nu,\lambda}=\{x\ :\ \tilde W (x) \geq 0\}$; otherwise,
either there would exist some $(x,y)$, with $x \in \tilde
\Omega_{\nu,\lambda}$ and $|y|>1$ such that $H(x,y)=H(y,x)=0$,
which is impossible from the analysis subsequent to
Equation~(3.1) in Section~3 of~\cite{CCG}, or there would be an equality $Y_e+(x) = Y_+(x) = 0$ at some $x \in [-1,1]\setminus \{0\}$, which is again impossible. Indeed, the only candidate for such a zero of $Y_+$ is $x=x_*$, which exists if $\nu^2 + \lambda^2 \neq 1$, and reads (see subsection 5.4 of~\cite{CCG}):
$$
x_* = \frac{-2\nu \lambda}{\nu^2 + \lambda^2 -1}\,.
$$
But then, it is easy to check that $Y_{e+}(x_*)=Y_+(\nu
x_*)/\nu \neq 0$, from the expression of $y_\pm$ written in
subsection 5.4 of~\cite{CCG}.
 \qed

We now have the main statement:
\begin{Theorem}
 The ergosurface always has a connected component, diffeomorphic to $\R\times S^1\times S^2$, \
lying beyond all Killing horizons,
     and an ``outer part", lying above the event horizon, such
     that:
\begin{itemize}
\item For $(\nu,\lambda)\in \mcU$ such that
    $\nu+\lambda<1$,  the outer ergosurface is
    diffeomorphic to $\R\times S^1\times S^2$.
\item For $(\nu,\lambda)\in \mcU$ such that
    $\nu+\lambda>1$, the outer ergosurface   is
    diffeomorphic to $\R\times \left(S^3\cup S^3\right)$,
    that is to say the space cross-sections have the
    topology of two disjoints copies of a 3-sphere.
\item In the limiting case $\nu+\lambda=1$, the outer
    ergosurface is a ``pinched'' $\R\times S^1\times S^2$.
\end{itemize}
\label{theo:ersurface-topology}
\end{Theorem}
\proof  From the previous results in this section, in particular from Proposition~\ref{P21I.1-}, the inner part of the ergosurface
is defined for all allowed values of the parameters, lies under the Killing horizon $\{y=y_c\}$, ``under'' the singular set, and meets the latter only at $(x,Y)=(0,0)$. In particular,
it is smooth, connected, reaches the axis of rotation $x=\pm 1$ related to the coordinate $\varphi$, and is therefore diffeomorphic to  $\R\times S^1\times S^2$.

For the outer part, we study its intersection with the set $\{y=-1\}$. To do so, recall the formula:
$$H(-1,x) = (1 + \nu - \lambda)\left(\nu(1 + \lambda - \nu)x^2 + 1 - \nu - \lambda\right)\ .$$
In particular, we have $H(-1,\pm 1) = (1 - \nu)(1 + \nu - \lambda)^2 > 0$. Since $H(y_h,\pm 1) < 0$, we obtain that $y_{e-}(\pm 1)$ are both in $(y_h,-1)$.

Then, the roots $x$ of $H(-1,x)=0$ exist if and only if $\nu + \lambda \geq 1$, and read
$$x \in \left\{\pm \sqrt{\frac{\nu + \lambda -1}{\nu(1 + \lambda - \nu)}}\right\}\ .$$
Those values of the roots are in $(-1,1)$ for all allowed values of $\nu$ and $\lambda$ such that $\nu + \lambda \geq 1$, as shown in the proof of Lemma 3.3 in Section 3 of~\cite{CCG}.

From this, we conclude that if $\nu + \lambda < 1$, the whole graph of $y_{e-}$ forms a connected set in the spacetime,
above the event horizon, and since it is smooth and defined for all allowed values of the parameter $x$,
it is again diffeomorphic to $\R\times S^1\times S^2$.

If $\nu + \lambda > 1$, the intersection of the graph of
$y_{e-}$ with the spacetime has two connected components,
$\{y_{e-}(x)\ :\ x \in [-1,- \sqrt{\frac{\nu + \lambda
-1}{\nu(1 + \lambda - \nu)}}]\}$ and $\{y_{e-}(x)\ :\ x \in
[\sqrt{\frac{\nu + \lambda -1}{\nu(1 + \lambda - \nu)}},1]\}$.
Therefore, after taking into account the rotations along the
coordinates $\varphi$ and $\psi$, we obtain that the part of
the outer ergosurface which lies in the spacetime is
diffeomorphic to $\R \times (S^3 \cup S^3)$ (the product of the
real line with two disjoint $3$-spheres). In particular, it has
two connected components.

In the limiting case $\nu + \lambda = 1$, the graph of $y_{e-}$
intersects the set $\{y=-1\}$ exactly once, at $x=0$.
Otherwise, it is located in the spacetime. The ergosurface in
this case is therefore diffeomorphic to a ``pinched" $\R \times
S^1\times S^2$.
\qed

\section{Acknowledgements}
The author is grateful to
P.T. Chru\'{s}ciel and to A. Garc\'ia-Parrado for many valuable
suggestions and comments.


\begin{thebibliography}{5}

\bibitem{ChCo}
P.T. Chru\'{s}ciel and J.~Cortier, \emph{{Maximal analytic extensions of the {Emparan-Reall}
  black ring}},  J. Diff. Geom. \textbf{85} (2010) 425-459.


\bibitem{CCG}
P.T. Chru\'{s}ciel, J.~Cortier and A. Garc\'ia-Parrado G\'omez-Lobo, \emph{On the global structure of
the Pomeransky-Senkov black holes}, (2009), arXiv:0911.0802 [gr-qc].


\bibitem{Durkee}
M.~Durkee, \emph{Geodesics and symmetries of doubly spinning black rings},
  Class.\ Quantum Grav. \textbf{26} (2009), 085016, 33. 

\bibitem{EmparanReall}
R.~Emparan and H.S. Reall, \emph{A rotating black ring in five dimensions},
  Phys.\ Rev.\ Lett. \textbf{88} (2002), 101101, arXiv:hep-th/0110260.

\bibitem{PS}
A.A. Pomeransky and R.A. Sen'kov, \emph{{Black ring with two angular momenta}},
   (2006), hep-th/0612005.


\end{thebibliography}
\end{document}